\documentclass[12pt]{article}
\begin{document}

\title{\bf Teleparallel Versions of Friedmann and Lewis-Papapetrou Spacetimes}

\author{M. Sharif \thanks{msharif@math.pu.edu.pk} and M. Jamil Amir
\thanks{mjamil$\_$dgk@yahoo.com}\\
Department of Mathematics, University of the Punjab,\\
Quaid-e-Azam Campus, Lahore-54590, Pakistan.}

\date{}

\maketitle

\begin{abstract}
This paper is devoted to investigate the teleparallel versions of
the Friedmann models as well as the Lewis-Papapetrou solution. We
obtain the tetrad and the torsion fields for both the spacetimes. It
is shown that the axial-vector vanishes for the Friedmann models. We
discuss the different possibilities of the axial-vector depending on
the arbitrary functions $\omega$ and $\psi$ in the Lewis-Papapetrou
metric. The vector related with spin has also been evaluated.
\end{abstract}


\section{Introduction}

The dynamics of the gravitational field can be described with the
help of Teleparallel theory (TPT) [1]. This theory is
characterized by the vanishing of curvature identically but the
torsion is taken to be non-zero. The basic entity of this theory
is the non-trivial tetrad field ${h^a}_\mu$ while in General
Relativity (GR) the metric tensor plays the role of the basic
entity. TPT corresponds to a gauge theory for the translation
group [2,3] based on Weitzenb$\ddot{o}$ck geometry [4]. In spite
of these fundamental differences, the two theories provide
equivalent descriptions of the gravitational interaction [5]. This
implies that curvature and torsion might be simply alternative
ways of describing the gravitational field. Consequently, these
are related to the same degrees of freedom of gravity. This
supports the fact that the symmetric energy-momentum tensor is a
source in both the theories, i.e., the source of curvature in GR
and the source of torsion in TPT. In some other theories [2,6],
torsion is only relevant when spins are important [7]. This point
of view indicates that torsion might represent additional degrees
of freedom as compared to curvature. As a result, some new physics
may be associated with it.

Even if GR is the unique true theory of gravity, consideration of
close alternative models can shed light on the properties of GR
itself. Theories of gravity based on the geometry of distance
parallelism [3, 8-14] are commonly considered as the closest
alternative to the GR. TP gravity models possess a number of
attractive features both from the geometrical and physical
viewpoints. Teleparallelism is naturally formulated by gauging
external (spacetime) translation and underline the
Weitzenb$\ddot{o}$ck spacetime characterized by the metricity
condition and by the vanishing of the curvature tensor. Translations
are closely related to the group of general coordinate
transformations which underlies GR. Thus the energy-momentum tensor
represents the matter source in the field equations of tetradic
theories of gravity like in GR.

There is a literature available [15-21] about the study of TP
versions of the exact solutions of GR. Recently, Pereira, at el.
[22] obtained the TP versions of the Schwarzschild and the
stationary axisymmetric Kerr solutions of GR. They proved that the
axial-vector torsion plays the role of the gravitomagnetic component
of the gravitational field in the case of slow rotation and weak
field approximations. Also, Nashed [23-25] has found many TP
versions of the exact solutions in GR and used them to calculate
different quantities. In this paper, we extend the procedure to find
the TP versions of the Friedmann models and the stationary
axisymmetric Lewis-Papapetrou solution of GR. It turns out that the
axial-vector has only two non-vanishing components for the
Lewis-Papapetrou spacetime as expected. However, the axial-vector
torsion vanishes for the Friedmann models due to spherical symmetry.
This is similar to the case of Schwarzschild spacetime [22].

The structure of the paper is as follows. In section $2$, we shall
briefly review the main results of the teleparallel theory. Section
$3$ is devoted to determine the tetrad field, the
Weitzenb$\ddot{o}$ck connection and the irreducible components of
the torsion tensor for the Friedmann models. Section $4$ provides
the evaluation of the tetrad field, the Weitzenb$\ddot{o}$ck
connection and the irreducible components of the torsion tensor for
the Lewis-Papapetrou spacetime. These will give the vector and the
axial-vector parts of the torsion tensor. We shall summarize and
conclude the results in the last section.

\section{An Overview of the Teleparallel Theory}

We define the Weitzenb$\ddot{o}$ck connection as [26]
\begin{eqnarray}
{\Gamma^\theta}_{\mu\nu}={{h_a}^\theta}\partial_\nu{h^a}_\mu,
\end{eqnarray}
where the non-trivial tetrad ${h^a}_\mu$ with its inverse field
${h_a}^\nu$ satisfies the relations
\begin{eqnarray}
{h^a}_\mu{h_a}^\nu={\delta_\mu}^\nu; \quad\
{h^a}_\mu{h_b}^\mu={\delta^a}_b.
\end{eqnarray}
In this paper the Latin alphabet $(a,b,c,...=0,1,2,3)$ will be
used to denote the tangent space indices and the Greek alphabet
$(\mu,\nu,\rho,...=0,1,2,3)$ to denote the spacetime indices. The
Riemannian metric in TPT arises as a by product [3] of the tetrad
field given by
\begin{equation}
g_{\mu\nu}=\eta_{ab}{h^a}_\mu{h^b}_\nu,
\end{equation}
where $\eta_{ab}$ is the Minkowski spacetime such that
$\eta_{ab}=diag(+1,-1,-1,-1)$. In TPT, the gravitation is attributed
to torsion [20], which plays the role of force. For the
Weitzenb$\ddot{o}$ck spacetime, the torsion is defined as [27]
\begin{equation}
{T^\theta}_{\mu\nu}={\Gamma^\theta}_{\nu\mu}-{\Gamma^\theta}_{\mu\nu},
\end{equation}
which is antisymmetric in nature. Due to the requirement of
absolute parallelism the curvature of the Weitzenb$\ddot{o}$ck
connection vanishes identically. The Weitzenb$\ddot{o}$ck
connection also satisfies the relation given by
\begin{equation}
{{\Gamma^{0}}^\theta}_{\mu\nu}={\Gamma^\theta}_{\mu\nu}
-{K^{\theta}}_{\mu\nu},
\end{equation}
where
\begin{equation}
{K^\theta}_{\mu\nu}=\frac{1}{2}[{{T_\mu}^\theta}_\nu+{{T_\nu}^
\theta}_\mu-{T^\theta}_{\mu\nu}]
\end{equation}
is the {\bf contortion tensor} and
${{\Gamma^{0}}^\theta}_{\mu\nu} $ are the Christoffel symbols of
GR given by
\begin{equation}
{{\Gamma^{0}}^\theta}_{\mu\nu}={\frac{1}{2}}g^{\theta\rho}(g_{\mu\rho,
\nu} +g_{\nu\rho, \mu}-g_{\mu\nu, \rho}).
\end{equation}
The torsion tensor of the Weitzenb$\ddot{o}$ck connection can be
decomposed into three irreducible parts under the group of global
Lorentz transformations [3]: the tensor part
\begin{equation}
t_{\lambda\mu\nu}={\frac{1}{2}}(T_{\lambda\mu\nu}
+T_{\mu\lambda\nu})+{\frac{1}{6}}(g_{\nu\lambda}V_\mu
+g_{\nu\mu}V_\lambda)-{\frac{1}{3}}g_{\lambda\mu}V_\nu,
\end{equation}
the vector part
\begin{equation}
{V_\mu}={T^\nu}_{\nu\mu},
\end{equation}
and the axial-vector part
\begin{equation}
{A^\mu}=\frac{1}{6}\epsilon^{\mu\nu\rho\sigma} T_{\nu\rho\sigma}.
\end{equation}
The torsion tensor can now be expressed in terms of these
irreducible components as follows:
\begin{equation}
T_{\lambda\mu\nu}={\frac{1}{2}}(t_{\lambda\mu\nu}-t_{\lambda\nu\mu})
+{\frac{1}{3}}(g_{\lambda\mu}V_\nu -
g_{\lambda\nu}V_\mu)\epsilon_{\lambda\mu\nu\rho}A^\rho,
\end{equation}
where
\begin{equation}
\epsilon^{\lambda\mu\nu\rho}= \frac{1}{\surd{-g}}
\delta^{\lambda\mu\nu\rho}.
\end{equation}
Here $\delta=\{\delta^{\lambda\mu\nu\rho}\}$ and
$\delta^*=\{\delta_{\lambda\mu\nu\rho}\}$ are completely skew
symmetric tensor densities of weight -1 and +1 respectively [3]. TPT
provides an alternate description of Einstein's equations which is
given by the teleparallel equivalent of GR [10, 26-27]. It is worth
mentioning here that the deviation of axial symmetry from the
spherical symmetry is represented by the axial-vector torsion. It
has been shown, in both GR and TPT, by many authors [3, 28] that the
spin precession of a Dirac particle in torsion gravity is related to
the torsion axial-vector by
\begin{equation}
\frac{d\textbf{S}}{dt}=- \frac{3}{2}\textbf{A}\times \textbf{S},
\end{equation}
where $\textbf{S}$ is the spin vector of a Dirac particle and
$\textbf{A}$ is the spacelike part of the torsion axial-vector. The
Hamiltonian would be of the form [29]
\begin{equation}
\delta H =-\frac{3}{2} \textbf{A}.{\bf\sigma},
\end{equation}
where ${\bf\sigma}$ is the particle spin.

\section{Teleparallel Solution of the Friedmann Models}

The Friedmann models of the universe are defined by the metric
\begin{equation}
ds^2=dt^2-a^2(t)[d\chi^2+{f_\kappa}^2(\chi) d\Omega^2],
\end{equation}
where
\begin{eqnarray}
f(\chi)&=&\sinh\chi,\quad k=-1,\nonumber\\
&=&\chi,\quad\quad\quad k=0,\nonumber\\
&=&\sin\chi,\quad k=+1,
\end{eqnarray}
$\chi$ is the hyper-spherical angle and $a(t)$ is the scale
parameter, $d\Omega^2=d\theta^2+\sin^2{\theta}d\phi^2$ is the
solid angle.

We use an isotropic form of the Friedmann metric to find the
components of the tetrad field. The Friedmann metric in isotropic
coordinates $(\tau,\rho,\theta,\phi)$ can be written as [30]
\begin{equation}
ds^2=d\tau^2-{\frac{a^2(\tau)}{(1+{\frac{1}{4}}\kappa\rho^2)^2}}
(d\rho^2+\rho^2d\Omega^2),
\end{equation}
where $\tau$ denotes the proper time. The proper radius is given
by
\begin{equation}
R(\tau)=\frac{\rho}{a(\tau)}\{1+{\frac{1}{4}}\kappa\rho^2\}.
\end{equation}
For the sake of simplicity, we substitute $
A(\rho)={1+{\frac{1}{4}}\kappa\rho^2}$ in Eq.(17) so that
\begin{equation}
ds^2=d\tau^2-{\frac{a^2(\tau)}{A^2(\rho)}}
(d\rho^2+\rho^2d\Omega^2).
\end{equation}
The tetrad components of the Friedmann models can thus be evaluated
by using a standard procedure [22, 26]. They are given as
\begin{equation}
{h^a}_\mu=\left\lbrack\matrix { 1   &&&   0    &&&   0    &&&   0
\cr 0        &&& a/A &&&   0    &&&   0 \cr 0 &&&    0 &&& a/A &&&
0 \cr 0        &&&   0    &&& 0    &&& a/A \cr } \right\rbrack.
\end{equation}
Its inverse becomes
\begin{equation}
{h_a}^\mu=\left\lbrack\matrix { 1   &&   0    &&   0    &&   0 \cr
0        &&  A/a &&   0    &&   0 \cr 0 &&    0 && A/a && 0 \cr 0
&&   0    && 0    && A/a \cr } \right\rbrack.
\end{equation}
If we replace $dt^2$ by $\gamma^2d\tau^2$ in Eq.(15) and then
compare it with Eq.(19), we get
\begin{eqnarray}
\gamma=1, \quad\          d\rho=A(\rho)d\chi,\\
f_\kappa(\chi)=\frac{\rho}{A(\rho)}.
\end{eqnarray}
Using general coordinate transformation law
\begin{equation}
{h^a}_\mu\acute{}={\frac{\partial{X^\nu}}{\partial{X^\mu\acute{}}}}
{h^a}_\nu,
\end{equation}
it follows that
\begin{equation}
{h^a}_\mu\acute{}=\left\lbrack\matrix {1   &&&   0    &&&   0 &&&
0 \cr  0        &&& a sin\theta\cos\phi &&&  af cos\theta\cos\phi
&&&   -af sin\theta\sin\phi\cr 0        &&& a sin\theta\sin\phi
&&& af cos\theta\sin\phi &&&   af sin\theta\cos\phi \cr 0 &&&   a
cos\theta   &&&   -af sin\theta   &&& 0 \cr } \right\rbrack.
\end{equation}
Its inverse is
\begin{equation}
{h_a}^\mu\acute{}=\left\lbrack\matrix { 1   &&&   0    &&&   0
&&&   0 \cr  0 &&& a^{-1} sin\theta\cos\phi
&&&(af)^{-1}cos\theta\cos\phi &&&   -(af sin\theta)^{-1}sin\phi
\cr  0        &&&   a^{-1} sin\theta\sin\phi   &&&
(af)^{-1}cos\theta\sin\phi &&&   (af sin\theta)^{-1}cos\phi\cr
0    &&&  a^{-1} cos\theta  &&&  -(af)^{-1}sin\theta    &&&  0
\cr } \right\rbrack.
\end{equation}
Notice that Eqs.(2) and (3) can be verified by using Eqs.(25) and
(26). When we use Eqs.(25) and Eq.(26) in Eq.(1), the following
non-vanishing components of the Weitzenb$\ddot{o}$ck connection turn
out:
\begin{eqnarray}
{\Gamma^1}_{10}&=&{\Gamma^2}_{20}={\Gamma^3}_{30}={\frac{\dot{a}}{a}},\nonumber\\
{\Gamma^1}_{22}&=&-f_\kappa(\chi),\quad\
{\Gamma^1}_{33}={\Gamma^1}_{22}sin^2\theta,\nonumber\\
{\Gamma^2}_{12}&=&{\Gamma^3}_{13}={\frac{1}{f_\kappa(\chi)}},
\quad\ {\Gamma^2}_{21}={\Gamma^3}_{31}={\Gamma^2}_{12}
f'_\kappa(\chi) ,\nonumber\\
{\Gamma^2}_{33}&=&-\sin{\theta}\cos\theta, \quad\
{\Gamma^3}_{23}=\cot\theta={\Gamma^3}_{32},
\end{eqnarray}
where dot and prime denote the derivatives w.r.t. $t$ and $\chi$
respectively. The corresponding non-vanishing components of the
torsion tensor are obtained by using Eq.(27) in Eq.(4). These are
given by
\begin{eqnarray}
{T^1}_{10}&=&{T^2}_{20}={T^3}_{30}=- {T^1}_{01}=- {T^2}_{02}=-
{T^3}_{03}=-\frac{\dot{a}}{a},\nonumber\\
{T^2}_{21}&=&{T^3}_{31}=-{T^2}_{12}=
-{T^3}_{13}=\frac{1}{f_\kappa(\chi)}\{1-{f'}_\kappa(\chi)\}.
\end{eqnarray}
Since $A^\mu$ gives the deviation from spherical symmetry [28], the
axial-vector part of the torsion tensor vanishes identically for the
Friedmann models, i.e,
\begin{equation}
A^\mu=0.
\end{equation}
One can verify this by using Eq.(28) in Eq.(10). This shows that the
spin vector of the Dirac particle is constant and the corresponding
Hamiltonian induced by the axial vector spin coupling vanishes. The
non-zero components of the vector and tensor parts of the torsion
tensor take the following forms:
\begin{eqnarray}
V_0&=&-\frac{3\dot{a}}{a},\\
V_1&=&\frac{2}{f_\kappa(\chi)}\{1-{f'}_\kappa(\chi)\}.
\end{eqnarray}
and
\begin{eqnarray}
t_{202}&=&-{\frac{1}{2}}a\dot{a}{f^2}_\kappa(\chi),\\
t_{220}&=&-2t_{202},\\
t_{303}&=&t_{202}sin^2\theta,\\
t_{330}&=&-2t_{303},\\
t_{212}&=&\frac{1}{2}a^2f_\kappa(\chi)\{1-f'_\kappa(\chi)\},\\
t_{221}&=&-2t_{212},\\
t_{313}&=&\frac{1}{2}a^2 f_\kappa(\chi)\{1-f'_\kappa(\chi)\}sin^2\theta,\\
t_{331}&=&-2 t_{313},
\end{eqnarray}
respectively. Since the torsion plays the role of the gravitational
force in TPT, a spinless particle will obey the force equation [22,
26] in the gravitational field
\begin{equation}
\frac{du_\rho}{ds}-\Gamma_{\mu\rho\nu} u^\mu u^\nu
=T_{\mu\rho\nu} u^\mu u^\nu.
\end{equation}
The left hand side of this equation is the Weitzenb$\ddot{o}$ck
covariant derivative of $u_\rho$ along the world line of the
particle. The presence of the torsion tensor on its right hand side
means essentially that torsion plays the role of an external force
in teleparallel gravity.

\section{Teleparallel Solution of the Lewis-Papapetrou Spacetime}

The class of stationary axisymmetric  solutions of the Einstein
field equations is the appropriate framework to include the
gravitational effect of an {\it external} source in an exact
analytic manner [31]. At the same time, such spacetimes are of
obvious astrophysical importance, as they describe the exterior of
the body in equilibrium. The line element of a stationary
axisymmetric spacetime is given by the Lewis-Papapetrou metric as
\begin{equation}
ds^2=e^{2\psi}(dt-\omega d\theta)^2-
e^{2(\gamma-\psi)}(d\rho^2+dz^2)- \rho^2e^{-2\psi}d\theta^2.
\end{equation}
Here $\omega$ is the angular velocity and $\gamma, \psi, \omega $
are arbitrary functions of $\rho$ and $z$ only. The corresponding
tetrad components are
\begin{equation}
{h^a}_\mu=\left\lbrack\matrix {e^\psi &&& 0 &&& -\omega e^\psi &&&
0 \cr 0 &&& e^{\gamma-\psi}cos\theta &&& -\rho e^{-\psi} sin\theta
&&& 0 \cr 0 &&& e^{\gamma-\psi} \sin \theta &&& \rho e^{-\psi}
cos\theta &&& 0 \cr 0 &&& 0 &&& 0 &&& e^{\gamma-\psi} \cr }
\right\rbrack.
\end{equation}
Its inverse is given by
\begin{equation}
{h_a}^\mu=\left\lbrack\matrix {e^{-\psi} && 0 && 0 && 0 \cr
-\omega\rho^{-1}e^{\psi} sin\theta  && e^{-\gamma+\psi} cos\theta
&& -\rho^{-1} e^\psi sin\theta   &&   0 \cr \omega\rho^{-1}e^\psi
cos\theta && e^{-\gamma+\psi}sin\theta && \rho^{-1}e^{\psi}
cos\theta && 0 \cr 0        &&   0   &&  0   && e^{-\gamma+\psi}
\cr } \right\rbrack.
\end{equation}
We see that Eqs.(2) and (3) can be easily verified by using Eqs.(42)
and (43). Using Eqs.(42) and (43) in Eq.(1), we obtain the following
non-vanishing components of the Weitzenb$\ddot{o}$ck connection
\begin{eqnarray}
{\Gamma^0}_{01}&=&\dot{\psi},\quad\ {\Gamma^0}_{03}=\psi',
\quad\ {\Gamma^0}_{12}=\omega \rho^{-1} e^\gamma, \nonumber\\
{\Gamma^0}_{21}&=&\omega \rho^{-1}-(\dot{\omega}
+2\omega\dot{\psi}), \quad\, {\Gamma^0}_{23}=-(\omega'+2\omega
\psi'),\nonumber\\
{\Gamma^1}_{11}&=&\dot{\gamma}-\dot{\psi}={\Gamma^3}_{31},\quad\
{\Gamma^1}_{22}=-\rho e^{-\gamma}, \quad\,
{\Gamma^1}_{13}=\gamma' -\psi'={\Gamma^3}_{33}, \nonumber\\
{\Gamma^2}_{12}&=&\rho^{-1} e^\gamma, \quad
{\Gamma^2}_{21}=\rho^{-1}(1-\rho\dot{\psi}),\quad\
{\Gamma^2}_{23}=-\psi' ,
\end{eqnarray}
where dot and prime denote the derivatives w.r.t. $\rho$ and $z$
respectively. The corresponding non-vanishing components of the
torsion tensor are obtained by using Eq.(44) in Eq.(4). These are
given by
\begin{eqnarray}
{T^0}_{01}&=&-\dot{\psi}=-{T^0}_{10}, \quad\
{T^0}_{03}=-\psi'=-{T^0}_{30},\nonumber\\
{T^0}_{12}&=&\omega\rho^{-1}(1-e^\gamma)-(\dot{\omega}+
2\omega\dot{\psi})=-{T^0}_{21},\nonumber\\
{T^0}_{23}&=&\omega'+2\omega\psi'= -{T^0}_{32},\quad
{T^1}_{13}=-\gamma'+\psi'=-{T^1}_{31}, \nonumber\\
{T^2}_{12}&=&\rho^{-1}(1-e^\gamma)-\dot{\psi} =-{T^2}_{21}, \quad
{T^2}_{23}=\psi'=-{T^2}_{32}, \nonumber\\
{T^3}_{31}&=&-\dot{\gamma}+\dot{\psi}=-{T^3}_{13}.
\end{eqnarray}
If we make use of Eq.(45) in Eq.(9), the following non-vanishing
components of the vector torsion turn out
\begin{eqnarray}
V_{1}&=&\dot{\psi}-\dot{\gamma}-\rho^{-1}(1-e^\gamma), \\
V_{3}&=&\psi'- \gamma'.
\end{eqnarray}
In view of Eq.(45), it follows from Eq.(10) that the non-vanishing
components of the axial-vector torsion are
\begin{eqnarray}
A^{(1)}&=&\frac{1}{3h}[ g_{00} {T^0}_{32}+g_{02}( {T^0}_{30}+ {T^2}_{32})], \\
A^{(3)}&=&\frac{1}{3h}[ g_{00} {T^0}_{12}+g_{02}
({T^2}_{12}+{T^0}_{01})],
\end{eqnarray}
where $h=\sqrt{-g}=\rho e^{(\gamma-\psi)}$. The component of axial
vector along $\theta$ direction vanishes due to symmetry about
${\rho}z$-plan. Therefore, the spacelike axial-vector can be written
as
\begin{equation}
\textbf{A}=\sqrt{-g_{11}}A^{(1)} \hat{e}_\rho+
\sqrt{-g_{33}}A^{(3)} \hat{e}_z,
\end{equation}
where $\hat{e}_\rho$ and $\hat{e}_z$ are the unit vectors along
radial and $z-$directions respectively. Using Eqs.(45)-(49) together
with the value of $h$ in Eq.(50), the axial-vector becomes
\begin{equation}
\textbf{A}=\frac{-1}{3\rho}e^{3\psi-\gamma}[(\omega'+
2\omega\psi')\hat{e}_\rho+\dot{\omega}\hat{e}_z].
\end{equation}
From Eq.(51), we note the following special cases depending upon the
values of $\omega$ and $\psi$. If $\omega$ is only a function of
$z$, then the axial-vector will be symmetric about radial axis and
takes the form
\begin{equation}
\textbf{A}=\frac{-1}{3\rho}e^{3\psi-\gamma}[(\omega'+
2\omega\psi')\hat{e}_\rho].
\end{equation}
If $\omega$ is only a function of $\rho$, then it lies in
${\rho}z$-plan and is given by
\begin{equation}
\textbf{A}=\frac{-1}{3\rho}e^{3\psi-\gamma}[
2\omega\psi'\hat{e}_\rho+\dot{\omega}\hat{e}_z].
\end{equation}
If $\omega$ is constant, $\textbf{A}$ becomes
\begin{equation}
\textbf{A}=\frac{-1}{3\rho}e^{3\psi-\gamma}
[2\omega\psi'\hat{e}_\rho],
\end{equation}
that is, it is symmetric about radial direction. Finally, when
$\omega$ is constant and also $\psi$ is a function of $\rho$ only,
then the axial-vector vanishes, i.e,
\begin{equation}
\textbf{A}=0,
\end{equation}
that is, in this case the cylindrical symmetry will not be disturbed
even if $\omega$ is non-zero. The spin precession of the Dirac
particle in torsion gravity turns out to be
\begin{equation}
\frac{d\textbf{S}}{dt}=\frac{1}{2\rho}e^{3\psi-\gamma}[(\omega'+
2\omega\psi')\hat{e}_\rho+\dot{\omega}\hat{e}_z]\times \textbf{S}.
\end{equation}
The corresponding Hamiltonian will be
\begin{equation}
\delta H=\frac{1}{2\rho}e^{3\psi-\gamma}[(\omega'+
2\omega\psi')\hat{e}_\rho+\dot{\omega}\hat{e}_z].{\bf\sigma}.
\end{equation}

\section{Summary and Discussion}

GR is very successful in describing long distance phenomena.
However, this theory encounters serious difficulties on
microscopic distances. The Lagrangian structure of GR differs, in
principle, from the ordinary microscopic gauge theories. In
particular, a covariant conserved energy-momentum tensor for the
gravitational field can not be constructed in the framework of GR.
Consequently, the study of alternative models of gravity is
justified from the physical as well as from the mathematical point
of view.

This paper is devoted to obtain TP versions for the Friedmann models
and the stationary axisymmetric Lewis-Papapetrou solution. For this
purpose, a tetrad having four unknown functions is applied to the
field equation of the TP theory of gravity by using a coordinate
transformation. The associated metrics of tetrad for the Friedmann
and the Lewis-Papapetrou spacetimes are given by Eqs.(25) and (42)
respectively. For the Friedmann models, due to spherical symmetry,
the axial-vector torsion vanishes identically and there occurs no
deviation in the spherical symmetry of the spacetime. Consequently
there exists no inertia field with respect to Dirac particle and the
spin vector of a Dirac particle becomes constant. The only
non-vanishing components of the vector part are the time and the
radial one. It is mentioned here that for the Schwarzschild metric,
the axial vector is only in the radial direction [22]. The reason is
that the Friedmann models are non-static while Schwarzschild metric
is static.

For the teleparallel Lewis-Papapetrou solution, we obtain the vector
and the axial-vector parts of the torsion tensor. The vector parts
are in the radial and $z$-directions. This corresponds with the Kerr
metric [22] where we get the vector part in the radial and
$\theta$-directions. The axial-vector torsion turns out to be
symmetric about ${\rho}z$-plan as its component along
$\theta$-directions vanishes everywhere. The non-inertia force on
the Dirac particle can be represented as a rotation induced torsion
of spacetime. There arise three possibilities of the axial-vector
depending upon the nature of the values of $\omega$ and $\psi$. When
$\omega$ is a function of $z$ only or $\omega$ is constant, the
axial-vector will be along the radial direction, that is, it will be
symmetric about radial direction. When $\omega$ is a function of
$\rho$ only, the axial-vector has its components along the radial as
well as z-direction, i.e., the axial-vector lies in ${\rho}z$-plan.
For the third possibility when $\omega$ is constant and $\psi$ is a
function of $\rho$ only, the axial-vector vanishes identically. For
the cases when $\omega$ is constant, the non-inertia force on the
Dirac particle remains same everywhere in the space [32]. From the
spacetime geometry view, the torsion axial-vector represents the
deviation from the symmetry of the underlying spacetime which
corresponds to an inertia field with respect to Dirac particle as
expressed by the Eq.(56).

Finally, we would like to mention here that tetrad formalism itself
has some advantages. This comes mainly from its independence from
the equivalence principle and consequent suitability to the
discussion of quantum issues. Some classic solutions of the Einstein
field equations have already been translated into the teleparallel
language. This paper adds two more solutions. It is always enriching
to look at known things from another point of view, so that the
endeavor is in itself commendable.

The study of energy content of this tetrad is in progress [33].

\vspace{0.5cm}

{\bf Acknowledgment}

\vspace{0.5cm}

We acknowledge the enabling role of the Higher Education Commission
Islamabad, Pakistan, and appreciate its financial support through
the {\it Indigenous PhD 5000 Fellowship Program Batch-I}. We would
like to thank anonymous referee for their useful suggestions.
\vspace{0.5cm}

{\bf References}

\begin{description}

\item{[1]} Maluf, J.W., da Rocha-Neto, J.F., Toribio, T.N.L. and
Castello-Branco, K.H.: Phys. Rev. {\bf D65}(2002)124001.

\item{[2]} Hehl, F.W., McCrea, J.D., Mielke, E.W. and Ne'emann, Y.: Phys.
Rep. {\bf 258}(1995)1.

\item{[3]} Hayashi, K. and Tshirafuji : Phys. Rev. {\bf D19}(1979)3524.

\item{[4]} Weitzenb$\ddot{o}$ck, R.: {\it Invarianten Theorie}
(Gronningen: Noordhoft, 1923).

\item{[5]} Gronwald, F. and Hehl, F.W.: {\it Proceedings of the
School of Cosmology and Gravitation on Quantum Gravity}, Eric,
Italy ed. Bergmann, P.G. et al. (World Scientific, 1995);\\
Blagojecvic, M. {\it Gravitation and Gauge Symmetries} (IOP
publishing, 2002).

\item{[6]} Hammond, R.T.: Rep. Prog. Phys. {\bf 65}(2002)599.

\item{[7]} Gronwald, F. and Hehl, F.W.: {\it On the Gauge Aspects of Gravity,
Proceedings of the 14th School of Cosmology and Gravitation},
Eric, Italy ed. Bergmann, P.G. et al. (World Scientific, 1996).

\item{[8]} Obukhov, Yu.N. and Pereira, J.G.: Phys. Rev. {\bf D64}(2002)027502.

\item{[9]} Hayashi, K.  and Nakano, T.: Prog. Theor. Phys. {\bf 38}(1967)491.

\item{[10]} De Andrade, V.L. and Pereira,  J.G.: Phys. Rev. {\bf
D56}(1997)4689.

\item{[11]} De Andrade, V.L, Guillen, L.C.T and Pereira, J.G.: Phys. Rev. Lett.
{\bf 84}(2000)4533.

\item{[12]} De Andrade, V.L, Guillen, L.C.T and Pereira, J.G.: Phys. Rev.
{\bf D64}(2001)027502.

\item{[13]} Kopezy$n'$ski, W.: J. Phys. {\bf A15}(1982)493.

\item{[14]} Nester, J.M.: Class. Quantum Grav. {\bf 5}(1988)1003.

\item{[15]} Hehl, F.W. and Macias, A.: Int. J. Mod. Phys. {\bf D8}(1999)399.

\item{[16]} Obukhov, Yu N., Vlachynsky, E.J., Esser, W.,
Tresguerres, R. and Hehl, F.W.: Phys. Lett. {\bf A220}(1996)1.

\item{[17]} Baekler, P., Gurses, M., Hehl, F.W. and McCrea, J.D.: Phys.
Lett. {\bf A128}(1988)245.

\item{[18]} Vlachynsky, E.J. Esser, W.,
Tresguerres, R. and Hehl, F.W.: Class. Quant. Grav. {\bf
13}(1996)3253.

\item{[19]} Ho, J.K., Chern, D.C. and Nester, J.M.: Chin. J. Phys.
{\bf 35}(1997)640.

\item{[20]} Hehl, F.W., Lord, E.A. and Smally, L.L.: Gen. Rel. Grav. {\bf 13}(1981)1037.

\item{[21]} Kawa, T. and Toma, N.: Prog. Theor. Phys. {\bf 87}(1992)583.

\item{[22]} Pereira, J.G., Vergas, T. and Zhang, C.M.: Class. Quant. Grav.
{\bf 18}(2001)833.

\item{[23]} Nashed, Gamal G.L.: Phys. Rev. {\bf D66}(2002)064015.

\item{[24]} Nashed, Gamal G.L.: Gen. Rel. Grav. {\bf34}(2002)1047.

\item{[25]} Nashed, Gamal G.L.: Nuovo Cimento {bf B117}(2002)521.

\item{[26]} Aldrovendi, R. and Pereira, J.G.: {\it An Introduction to
Gravitation Theory} (preprint)

\item{[27]} Aldrovandi and Pereira, J.G.: {\it An Introduction to
Geometrical Physics} (World Scientific, 1995).

\item{[28]} Nitsch, J. and Hehl, F.W.: Phys. Lett. {\bf B90}(1980)98.

\item{[29]} Mashhoon, B: Class. Quantum Grav. {\bf 17}(200)2399.

\item{[30]} Govender, M. and Dadhich, N.: Phys. Lett. {\bf B538}(2002)233.

\item{[31]} Galtsov, D.V.: arXiv:gr-qc/9808002.

\item{[32]} Zhang, C.M.: Mod. Phys. Lett. {\bf A16} (2001)2319.

\item{[33]} Sharif, M. and Amir, M. Jamil: submitted for publication.

\end{description}
\end{document}